# Low temperature photoluminescence imaging and time-resolved spectroscopy of single CdS nanowires

L.V. Titova*, Thang Ba Hoang, H.E. Jackson, and L.M. Smith
*Dept of Physics, Univ. Cincinnati*
J.M. Yarrison-Rice
*Dept. of Physics, Miami University*
J.L. Lensch, L.J. Lauhon
*Materials Science, Northwestern University*

Time-resolved photoluminescence (PL) and micro-PL imaging were used to study single CdS nanowires at 10 K. The low-temperature PL of all CdS nanowires exhibit spectral features near energies associated with free and bound exciton transitions, with the transition energies and emission intensities varying along the length of the nanowire. In addition, several nanowires show spatially localized PL at lower energies which are associated with morphological irregularities in the nanowires. Time-resolved PL measurements indicate that exciton recombination in all CdS nanowires is dominated by non-radiative recombination at the surface of the nanowires.

*Author to whom the correspondence should be addressed. Electronic mail:

ltitova@physics.uc.edu

Semiconductor nanowires are promising candidates for bottom-up approaches to nanoscale visible wavelength photonic devices. Optically and electrically-driven lasers,[1,2] photodetectors[3] and waveguides[4] have been demonstrated using single CdS nanowires fabricated by catalyst-assisted vapor-liquid-solid (VLS) growth. Although progress in CdS nanowire synthesis and device fabrication has been rapid, detailed studies of the fundamental optical and electronic properties of the nanowires and their dependence on the nanowire morphology has not been carried out.

In this letter, we show that low-temperature micro-photoluminescence (micro-PL) imaging combined with time-resolved PL (TRPL) measurements of *single* CdS nanowires can provide important information on the nature of electronic states and recombination modes in these structures. Previous cathodoluminescence measurements of single CdSe needles,[5] ZnSe nanorods[6] and ZnO nanowires[7] have indicated that morphology can influence their emission. Here, direct optical imaging of a number of single VLS-prepared CdS nanowires reveal two distinct types of emission: a spectral band emitting at energies associated with free and bound exciton emission from CdS, and spatially localized states at lower energy which we relate to morphological irregularities of the nanowires. Recombination dynamics show that the near-band edge emission (NBE) is dominated by non-radiative surface recombination, while the spatially localized states exhibit much longer recombination lifetimes, comparable to lifetimes seen for excitons in bulk CdS. For optimization of nanowire electro-optical devices this information implies that the quantum efficiency for emission *and* detection could be improved by an order of magnitude through passivation of the surface states. In addition, these results demonstrate that low-temperature PL measurements provide a rapid and nondestructive means for quantifying and selecting high quality single nanowires for further processing.

CdS nanowires were synthesized using techniques described previously[8] and deposited onto a silicon substrate via a dilute solution so that individual nanowires could be studied. The CdS nanowires under study ranged in diameter from 50 to 200 nm with lengths of 10 to 15 μm. For TRPL measurements, single CdS NWs were excited with 432 nm 200 fs pulses and the PL was detected with 80 ps resolution using a fast microchannel photomultiplier tube combined with time-correlated single photon counting. Spatially-resolved PL from single CdS nanowires was obtained through slit-confocal microscopy using the 458 nm line of a CW argon laser for excitation. Using our imaging technique described elsewhere,[9] we obtain a single CCD image which contains both spectral (with 70 μeV resolution) as well as spatial (with 1.2 μm resolution) information from a single nanowire. All measurements were carried out at 10 K.

TRPL measurements can provide insights into the nature of electronic states in nanowires and reveal dominant recombination mechanisms. Low-temperature band edge PL of bulk CdS, for instance, displays bound exciton complex emission, or free exciton emission, with lifetimes ranging from 500 to 1000 ps, reflecting the different electron-hole overlaps for different excitonic states.[10,11] Since the size of the nanowires studied here is large compared to the exciton Bohr radius in CdS (2.8 nm), quantum confinement effects are not significant, and one would expect low temperature nanowire emission to exhibit similar properties to the bulk, excepting properties that are sensitive to the presence of a surface. TRPL and micro-PL measurements were made on ten different

CdS nanowires. Here we present a detailed discussion of two of these nanowires which were selected to demonstrate the range of observed behavior.

In Fig. 1 (a) and (b) we show TRPL spectra from these two wires, (1 and 2) as false-color images (a log intensity scale) where the vertical axis corresponds to the time after the laser pulse, while the horizontal axis corresponds to the emission energy (with ~5 meV resolution). Wire 1 (Fig. 1 (a)) shows a single broad peak at an energy of 2.525 eV which decays rapidly. The energy of this peak is comparable to that of the excitonic emission from bulk CdS, but the measured exciton recombination lifetime (Fig. 1(c)) is only 80 ps (limited by our system response), nearly an order of magnitude shorter than observed in bulk CdS.

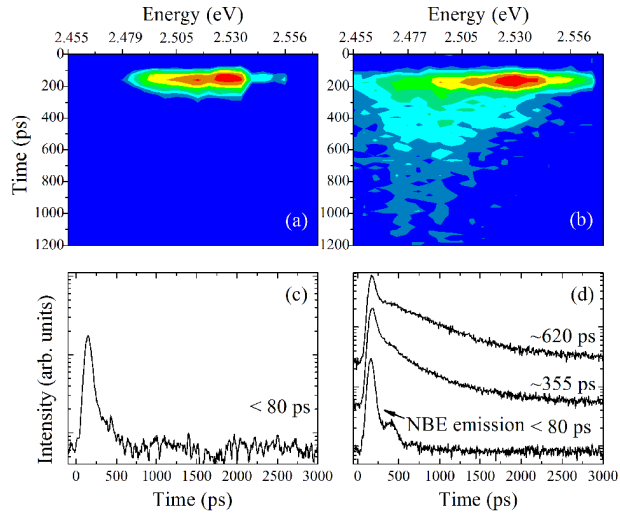

FIG. 1. (Color online) Time-resolved PL spectra from nanowires 1 (a) and 2 (b) at 10 K. Step size along the horizontal axis is 1 nm. Intensity (in logarithmic scale) is shown as a color scale. (c) Nanowire 1, decay taken at NBE emission energy, 2.525 eV. (d) Nanowire 2, Decays taken at NBE emission energy, 2.525 eV (bottom curve) and two energies (~2.472 eV and 2.486 eV) in the defect-related band (top and middle curves).

In contrast, while the TRPL spectrum from wire 2 (Fig. 1(b)) shows the same intense, but short-lived, NBE emission at early times, it also exhibits a series of sharp peaks at lower energy which persist to much longer times after the laser pulse. This can be seen in the corresponding time-decays displayed in Fig. 1(d). The recombination lifetime of the NBE emission at 2.525 eV from wire 2 is again only 80 ps. However, time-decays taken at two positions in the lower energy range (2.472 eV and 2.486 eV respectively) show distinctly longer recombination lifetimes of 355 and 620 ps.

The short (<80 ps) lifetime of the NBE emission displayed by all nanowires is likely to be caused by the close proximity of the non-radiative surface states due to the large surface-to-volume ratio of the nanowires. Optical emission of the strongly ionic materials such as CdS is known to be strongly affected by the dark trap states formed by the dangling bonds at the surfaces,[12] as well as by the surface states associated with any adsorbed impurities.[13] While the diameter of the nanowires is significantly larger than the exciton Bohr radius, the exciton diffusion length of around 1 μm at 7K[14] suggests that the excitons are likely to interact with the nanowire surfaces many times during their lifetimes, and are thus susceptible to annihilation at surface trap states.

To gain further insight into the origin of these two types of electronic states, we performed micro-PL imaging. A 2D PL map of wire 2 is displayed in Fig. 2 (a). In this image, the vertical axis denotes the spatial position along the nanowire, while the horizontal axis corresponds to the emission energy. An AFM image of wire 2 is displayed to the right of the PL image and shows that wire 2 is particularly non-uniform with many kinks, bends and lobes. The spectral map identifies clearly that the longer-lived lower energy emission observed in TRPL spectrum of wire 2 appears in the form of sharp spectral peaks, originating from specific localized points along the nanowire, superimposed on a broad background. Spectra (horizontal cross-sections of the CCD image) taken at four different positions (A, B, C and D) along the nanowire are shown in Fig. 2 (b). In this way, one can see that each sharp spectral feature is associated with a particular position along the wire. Moreover, the different emission energies of each line must also reflect differences in the local confining potentials. The NBE emission, while significantly less pronounced under CW excitation compared to the pulsed excitation, appears as a high energy shoulder in the emission spectrum.

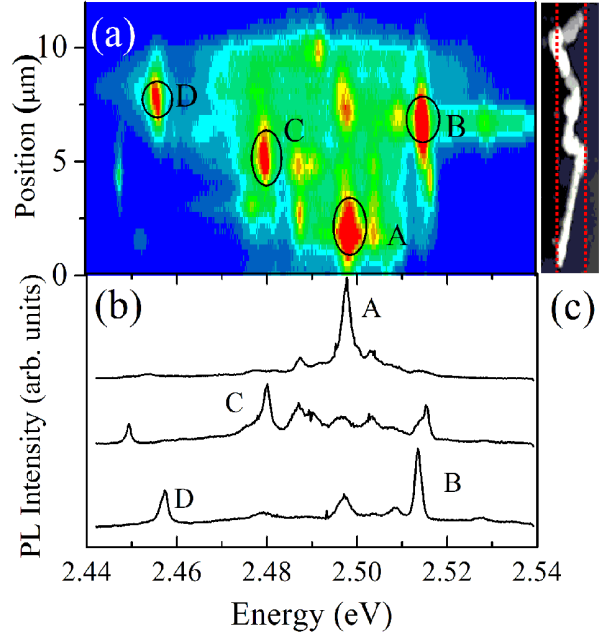

FIG. 2 (Color online) Photoluminescence imaging of the wire 2 at 10 K: (a) 2D map of the wire 1 emission; (b) PL spectra extracted from the 2D map at the positions specified by the circles in (a); (c) schematic description of the experiment - image of the spectrometer slit superimposed on the AFM micrograph of the wire.

In contrast to wire 2, the 2D PL image of wire 1 in Fig. 3 (a) shows no evidence of localized states. As can be seen in AFM image displayed to the right of Fig. 3, wire 1 is largely uniform and straight, and thus its spectra is dominated by the NBE broad emission peak. The intensity of the PL emission from this nanowire varies by almost an order of magnitude at different positions along the 14 μm length of the wire. More importantly, one can clearly see in the spatially resolved spectra extracted from the image in Fig. 3(b) that the PL peak position also varies significantly (by nearly 10 meV). The variations in emission intensity and energy are possibly related to non-uniform strain or compositional variation along the length of these nanowires.

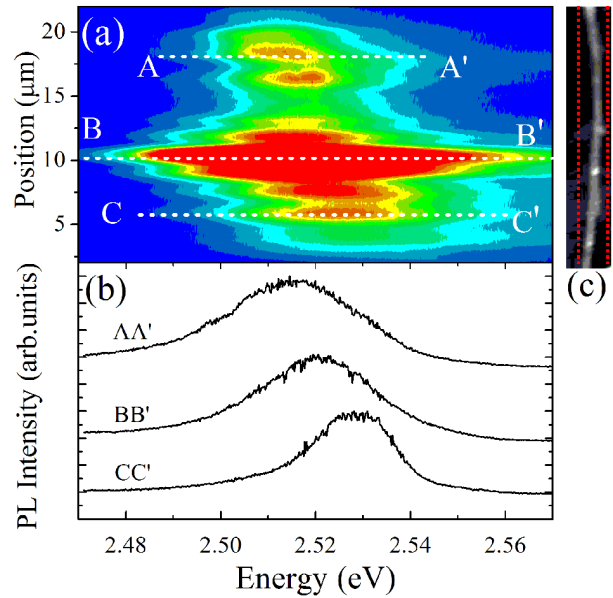

FIG. 3 (Color online) Photoluminescence imaging of the wire 1 at 10 K: (a) 2D map of the wire 1 emission; (b) PL spectra extracted from the 2D map at the positions specified by the lines in (a); (c) schematic description of the experiment - image of the spectrometer slit superimposed on the AFM micrograph of the wire.

From comparisons of nanowires of different structural quality, we conclude that emission of the longer-lived, spatially localized states which emit at lower energies result from structural imperfections. The emission energies of these states vary from line to line and do not precisely coincide with the known emission energies of excitons bound to bulk impurities. On the other hand, their recombination lifetimes, ranging from 300 ps to 1ns, are similar to those observed for bound exciton complexes in bulk CdS.[10,11] Thus we tentatively assign these states to excitons which are bound to structural defects and aggregates on the surface of the nanowires. Energies and electron-hole overlaps, and hence lifetimes of these states, are determined by the local confinement potential of the defect and therefore can vary significantly from position to position along the wire.

In summary, we have shown that low temperature time- and spatially-resolved PL measurements of single nanowires not only provide valuable information about the nature and properties of the electronic states, but also can be used as a sensitive measure of morphological quality. We find the PL spectra of irregular-shaped nanowires to exhibit localized states associated with the imperfections, with lifetimes comparable to exciton recombination observed in bulk CdS. The short lifetime ($< 80$ ps) of the band edge emission indicates significant influence of non-radiative surface states, suggesting that the quantum efficiency at low temperatures could be increased by over an order of magnitude by passivating the surface states through the growth of a core-shell structure or molecular passivation techniques.

**Acknowledgements:**

LVT and TBH are both equally responsible for this paper. This work was supported by the National Science Foundation through grants DMR 0071797 and 0216374, the Petroleum Research Fund of the American Chemical Society, the University of Cincinnati and Northwestern University. J.L.L. acknowledges the support of a National Science Foundation Graduate Research Fellowship.